%% file: main_final.tex
\documentclass[%
 reprint,
superscriptaddress,
 amsmath,amssymb,
 prl,
floatfix,
]{revtex4-2}
\pdfoutput=1

\usepackage{graphicx}
\usepackage{dcolumn}
\usepackage{bm}
\usepackage[colorlinks=true,
            linkcolor=blue,
            urlcolor=blue,
            citecolor=blue]{hyperref}
\usepackage[mathlines]{lineno}
\usepackage[all]{hypcap}
\usepackage{SIunits}
\usepackage{upgreek}
\usepackage[artemisia]{textgreek}
\usepackage{tabularx}
\usepackage{xcolor}
\usepackage[misc]{ifsym}

\makeatletter
\newcommand{\fmarki}{*}
\newcommand{\fmarkii}{\ensuremath{\dagger}}
\def\@fnsymbol#1{{\ifcase#1\or \fmarki\or \fmarkii\or \fmarkiii\or \fmarkiv\or \fmarkv\or \fmarkvi\or \fmarkvii\or \fmarkviii\or \fmarkix \else\@ctrerr\fi}}
\makeatother
\renewcommand{\fmarki}{\Letter}
\renewcommand{\fmarkii}{\Letter}

\newcommand{\beginsupplement}{%
        \setcounter{table}{0}
        \renewcommand{\thetable}{S\arabic{table}}%
        \setcounter{figure}{0}
        \renewcommand{\thefigure}{S\arabic{figure}}%
     }

\begin{document}

\title[Acoustic bursts during reconnection]{Ion and electron acoustic bursts during anti-parallel magnetic reconnection driven by lasers}

\author{Shu Zhang}
\email{shuzhang@princeton.edu}
 \affiliation{Department of Astrophysical Sciences, Princeton University, Princeton, New Jersey 08544, USA}
 
 \author{Abraham Chien}
 \affiliation{Department of Astrophysical Sciences, Princeton University, Princeton, New Jersey 08544, USA}
 
 \author{Lan Gao}
 \affiliation{Princeton Plasma Physics Laboratory, Princeton University, Princeton, New Jersey 08543, USA}
 
 \author{Hantao Ji }
\email{hji@pppl.gov}
 \affiliation{Department of Astrophysical Sciences, Princeton University, Princeton, New Jersey 08544, USA}
 \affiliation{Princeton Plasma Physics Laboratory, Princeton University, Princeton, New Jersey 08543, USA}
 
\author{Eric G. Blackman}
 \affiliation{Department of Physics and Astronomy, University of Rochester, Rochester, New York 14627, USA}
 
  \author{Russ Follett}
 \affiliation{Laboratory for Laser Energetics, University of Rochester, Rochester, New York 14623, USA}
 
  \author{Dustin~H.~Froula}
 \affiliation{Laboratory for Laser Energetics, University of Rochester, Rochester, New York 14623, USA}
 
   \author{Joseph Katz}
 \affiliation{Laboratory for Laser Energetics, University of Rochester, Rochester, New York 14623, USA}
 
 \author{Chikang Li}
 \affiliation{Plasma Science and Fusion Center, Massachusetts Institute of Technology, Cambridge, Massachusetts 02139, USA}
 
 \author{Andrew Birkel}
 \affiliation{Plasma Science and Fusion Center, Massachusetts Institute of Technology, Cambridge, Massachusetts 02139, USA}
 
 \author{Richard Petrasso}
 \affiliation{Plasma Science and Fusion Center, Massachusetts Institute of Technology, Cambridge, Massachusetts 02139, USA}
 
 \author{John Moody}
 \affiliation{Lawrence Livermore National Laboratory, Livermore, California 94550, USA}
 
 \author{Hui Chen}
 \affiliation{Lawrence Livermore National Laboratory, Livermore, California 94550, USA} 
 
\begin{abstract}
Magnetic reconnection converts magnetic energy into thermal and kinetic energy in plasma. Among numerous candidate mechanisms, ion acoustic instabilities driven by the relative drift between ions and electrons, or equivalently electric current, have been suggested to play a critical role in dissipating magnetic energy in collisionless plasmas. However, their existence and effectiveness during reconnection have not been well understood due to ion Landau damping and difficulties in resolving the Debye length scale in the laboratory. Here we report a sudden onset of ion acoustic bursts measured by collective Thomson scattering in the exhaust of anti-parallel magnetically driven reconnection using high-power lasers. The ion acoustic bursts are followed by electron acoustic bursts with electron heating and bulk acceleration. We reproduce these observations with 1D and 2D particle-in-cell simulations in which electron outflow jet drives ion-acoustic instabilities, forming double layers. These layers induce electron two-stream instabilities that generate electron acoustic bursts and energize electrons. Our results demonstrate the importance of ion and electron acoustic dynamics during reconnection when ion Landau damping is ineffective, a condition applicable to a range of astrophysical plasmas including near-Earth space, stellar flares, and black hole accretion engines.
\end{abstract}

\maketitle

Magnetic reconnection is a fundamental physical process through which energy is rapidly converted from magnetic field to plasma by alternating magnetic topology~\cite{yamada10,ji22}. 
Magnetic reconnection has been considered as a key energy release mechanism during solar and stellar flares~\cite{masuda94}, in Earth's magnetosphere~\cite{hesse20}, as well as during energetic phenomena in distant Universe such as the black hole’s accretion disk~\cite{dimatteo+1997,yuan14}.
It has been a longstanding challenge to identify the underlying kinetic mechanisms for efficient dissipation required for the topological change as well as energy conversion to explain the observed fast reconnection in nearly collisionless plasmas in space and astrophysics. There has been progress in understanding and confirming 2D kinetic mechanisms often represented by nongyrotropic pressure tensor~\cite{hesse99,kulsrud05,burch16,torbert18} in the electron diffusion regions where field lines break and reconnect. Beyond these 2D laminar processes, however, the kinetic dissipation mechanisms operating in general 3D are still much less understood~\cite{ji08,cozzani21} within or near diffusion regions and separatrices~\cite{lapenta15} that feature strong spatial gradients and streaming.  They include various plasma waves or instabilities, such as whistler waves~\cite{kennel66,goldman14}, Buneman instabilities~\cite{buneman58,drake03,che09}, lower-hybrid drift waves~\cite{carter02a,ji04,ji05,fox10,yoo18,graham19,chen20,yoo20} (due to cross-field gradient~\cite{krall71} or cross-field drift~\cite{mcbride72}), drift kink~\cite{daughton99} or kinetic Kelvin-Helmholtz~\cite{nakamura17} instabilities. 

Among these 3D candidate dissipation mechanisms, unstable ion-acoustic waves (IAWs)~\cite{coppi71,smith72,coroniti77,sagdeev79} driven by a relative drift between electrons and ions, or equivalently electric current, have attracted considerable interest as potential sources for the enhanced resistivity or viscosity that is often used within fluid descriptions as a local, current-dependent anomalous resistivity required for the sustained Petschek model of fast reconnection~\cite{ugai77,sato79,aparicio98,kulsrud98,kulsrud01,uzdensky03}. Despite early pioneering laboratory detection~\cite{gekelman84}, however, the importance of IAWs for magnetic reconnection has been quickly dismissed due to the widely observed high ion temperature ($T_{\rm i}/Z \gtrsim T_{\rm e}$) in space and in the laboratory where IAWs are strongly stabilized by ion Landau damping. Technical difficulties in the laboratory in measuring plasma waves in the short wavelengths on the order of Debye length have also prevented progress in identifying IAWs and understanding their detailed properties and role in magnetic reconnection.

In this Article, we present a laboratory platform where reconnection is driven magnetically at low-$\beta$ by laser-powered capacitor coils~\cite{chien22} in high-Z plasmas where $T_{\rm i}/Z \ll T_{\rm e}$. The sudden onset of bursts of IAWs is successfully measured in the exhaust region by collective Thomson scattering diagnostics. The IAWs are followed by bursts of electron acoustic waves (EAWs) with electron heating and bulk acceleration. The corresponding Particle-in-Cell (PIC) simulations in 1D and 2D show that IAWs are destabilized by an electron exhaust jet where the relative drift between ions and electrons is large.  IAWs grow rapidly to form electrostatic double layers, which in turn accelerate electrons to drive two-stream instability generating bursts of EAWs while heating electrons. Our results demonstrate the importance of ion and electron acoustic dynamics causing bursty energy dissipation during magnetic reconnection when ion Landau damping is ineffective. Implications for the reconnection process in magnetically dominated plasmas during stellar flares and accretion onto black holes are discussed.

\section*{Reconnection platform with laser-driven capacitor coils}

\begin{figure}[tb]
    \centering
    \includegraphics{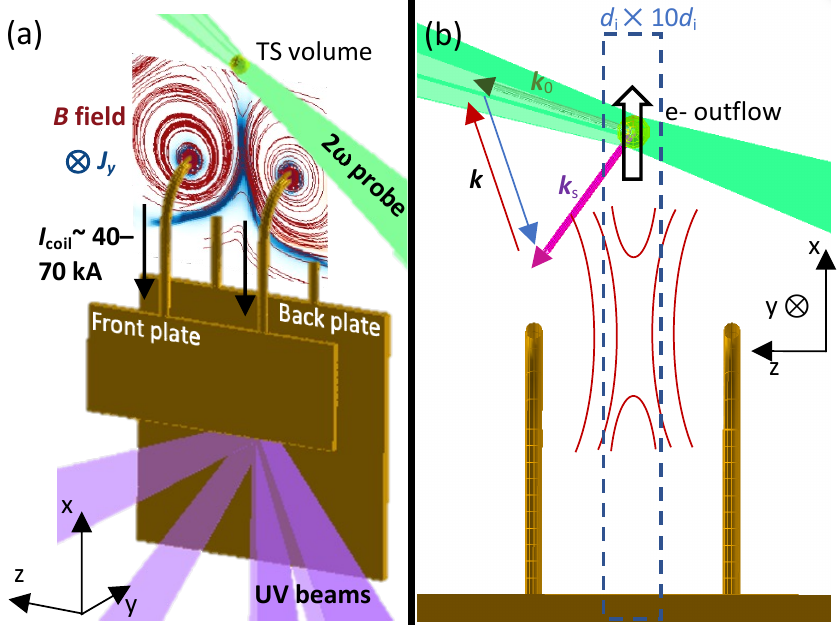} 
    \caption{Experimental setup and Thomson scattering diagnostics. 
    (a) Six UV beams (purple) are used to irradiate the back plate of the capacitor, driving current in the coils with $I_{\rm coil}\sim 40-70$~kA. Black arrows mark the current directions in the coils.  FLASH MHD simulation results are overlapped with the target to show the structure of the magnetic field (red lines) and the out-of-plane current density (blue) in the $y$-direction. 
    A 2$\omega$ (527 nm) Thomson scattering beam (green) probes the reconnection exhaust region, 600 µm above the center point between the top of the coils. The scattered light in a volume $60\times60\times50$~µm$^3$ is collected by an $f/10$ reflective collection system. (b) A face-on view of the reconnection region. The two vertical brown lines are the coils in (a). $\bm{k}_0$ and  $\bm{k}_{\rm s}$ are the wavevectors of the probe beam and the collected scattered light. The red and blue arrows indicate wavevectors ($\bm{k}$) of waves in plasma resonant with the probe and the scattered light. The red arrow is for the wave generating redshifted scattered light, and the blue arrow is for the wave generating blueshifted scattered light. These $\bm{k}$'s are in the $x-y$ plane and 17\degree{} off the outflow direction. The blue dashed box indicates a $d_{\rm i}\times 10d_{\rm i} = 180 ~\micro\meter{}\times 1800~\micro\meter{}$ region, in which ion skin depth $d_{\rm i}=c/\omega_{\rm pi}$, and $\omega_{\rm pi}$ is the Cu$^{18+}$ ion plasma frequency. Red lines illustrate the magnetic field lines, and the hollow arrow is the direction of the electron outflow jet measured by Thomson scattering in this experiment.
    }
    \label{fig:setup}
\end{figure}

The presented experiments were performed on the OMEGA laser facility at the Laboratory for Laser Energetics, University of Rochester~\cite{OMEGA}. The experimental platform is shown in Fig.~\ref{fig:setup}. The capacitor-coil target is made of a 50~\micro\meter{} thick Cu foil laser-cut to the shape of two plates connected by two wires. The wires, separated by 600 µm, are bent to 600 µm diameter half-circle coils and 500 µm straight legs. The target pictures are presented in supplemental Fig.~S4.
Six beams of 500~J 1-ns UV ($\lambda=351$~nm) lasers are focused on the center of the capacitor's back plate to drive a $\sim40-70$~kA current in the coils forming an anti-parallel magnetic field. The laser-generated plasmas diffuse into the region between the coils, and the x-rays and the current also heat the coils generating plasmas. The plasma between the coils is magnetized by the coil-driven antiparallel magnetic field, forming a reconnection current sheet. 

The magnetic field generated by the capacitor-coil targets is measured using proton radiography~\cite{li06prad}, similar to our previous experiments~\cite{gao16,chien19,chien21a, chien22}. The upstream magnetic field strength is $\sim23-40$~T at 6 ns after the lasers' onset. The proton radiographs also show the reconnection current sheet exists. 
The radiographs and analyses are presented in Supplementary.
To further quantify the reconnection conditions, we have conducted radiative and non-ideal magnetohydrodynamic (rad-MHD) simulations using the FLASH code~\cite{FLASH2000} to simulate the plasma diffusing from the capacitor plates and the plasma emerging from the heated coils due to Ohmic heating and x-ray radiation. The supplemental materials detail the setup of the non-ideal rad-MHD simulation. The simulated magnetic field lines and the current density at 3~ns, overlapped on the target in Fig.~\ref{fig:setup}(a), show that a reconnection current sheet is formed between the coils. This reconnection current sheet lasts until 10~ns, as shown in supplemental Fig.~S2. 
The synthetic proton radiographs shown in Fig.~S3(d,e) have a current sheet-induced central flask-like feature, consistent with the experimental one shown in Fig.~S3(a).

An approach to studying kinetic instabilities in reconnection is provided by combining a laser-driven capacitor-coil reconnection platform with collective Thomson scattering. Collective Thomson scattering diagnoses the spectrum of the density fluctuations in plasmas, which may be due to the microturbulence induced by kinetic instabilities~\cite{Milder2022, froula03,Daughney1970}. It can also diagnose the Stokes and anti-Stokes scattering of natural resonances in plasmas, such as ion-acoustic waves (IAW) and electron plasma waves (EPW). Since the plasma parameters determine the spectrum of the IAWs and EPWs-induced scattering, Thomson scattering is frequently used to diagnose the plasma's densities, temperatures, mean ion charge, flow speed, electron-ion relative drift as well as non-Maxwellian distributions~\cite{TSbook, suttle16,hare17, suttle21, sakai20, swadling20,  Milder2020}.

In this experiment, we recorded the Thomson scattering spectrogram of a probe laser ($\lambda=527$~nm, 150~J energy, 3.7~ns square pulse, 60~\micro\meter{} spot size) focused at 600~\micro\meter{} ($\sim3d_{\rm i}$) above the center between the top of the coils (reconnection x-line) as shown in Fig.~\ref{fig:setup}. As shown in Fig.~\ref{fig:setup}(b), the directions of the probe light and the scattered light collector determine the wavevector $\bm{k}$ of the measured density fluctuations or natural resonances since $\bm{k} =\bm{k}_{\rm 0}-\bm{k}_{\rm s}$ or $\bm{k} =\bm{k}_{\rm s}-\bm{k}_0$. $\bm{k}$ is in the direction 17$\degree$ from the outflow direction ($+\bm{x}$) and $k\sim k_0=2\pi/527~$nm$^{-1}$.
To infer the exhaust plasma's parameters, we fit the synthetic Thomson scattering spectra to the measured spectra. The synthetic spectrum is calculated based on Eq.~(\ref{eq:powerspectrum}) in the Method section. The least-squares fit suggests that the plasma in the exhaust region has electron density $n_{\rm e}\sim5\times10^{18}$~cm$^{-3}$, electron temperature $T_{\rm e}\sim 200-300$~eV, mean ion charge $Z\sim18$ and flow velocity $v\sim 1.5-3.5\times10^5$~m/s, which roughly matches the Alfven speed ($1.2-2.0\times10^5$~m/s).

\section*{Ion and electron acoustic bursts and electron heating}

\begin{figure*}[h]
    \centering
    \includegraphics{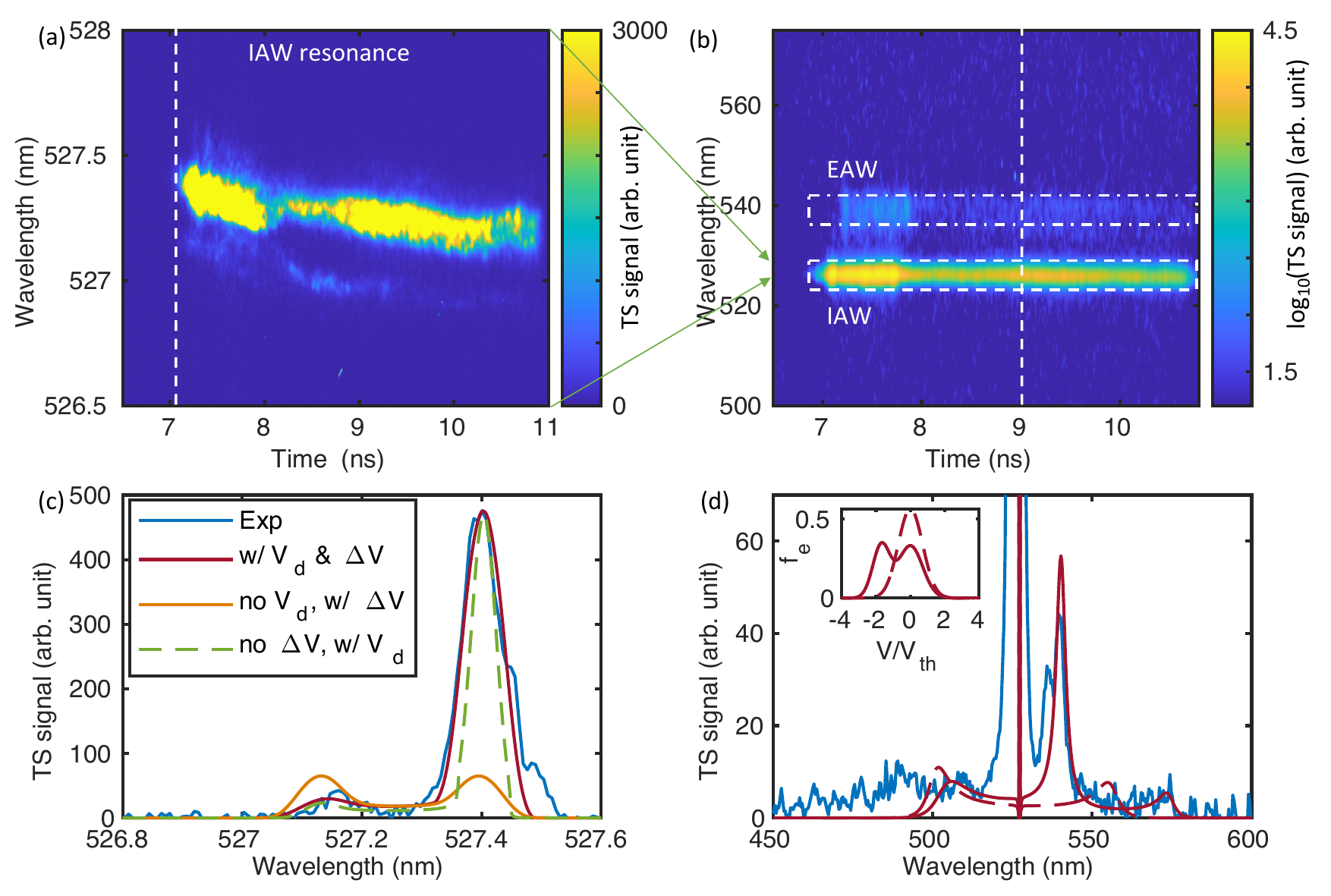} 
    \caption{Thomson scattering data and analysis. (a) narrowband (IAW) and (b) broadband time-resolved Thomson scattering (TS) at $t=7-10$~ns.  The IAW resonant peaks in (a) are highly asymmetric (grow from 10:1 to 100:1). The spectrum at 7.1 ns, before the IAW bursts, along the vertical dashed line in (a) is plotted in (c) as the blue line and compared with the synthetic TS spectra. The green dashed synthetic spectrum in (c) can reproduce the asymmetry of the IAW peaks, which is calculated with electrons streaming relative to ions with drift velocity $v_{\rm d}=0.17v_{\rm th}$ (electron thermal velocity) along the k-direction (red arrow) in Fig.~\ref{fig:setup}b.
    In addition to the electron drift, an inhomogeneous flow velocity with $\Delta v= 2\times10^{4}$~m/s$\sim$ $v_i$ (ion thermal velocity) in the scattering volume can broaden the IAW peak, shown as the red solid line, to match the experiment. A plasma without a drift but with an inhomogeneous flow velocity would generate a symmetric IAW spectrum, shown as the orange line. (b) is the spectrogram from the broadband spectrometer showing both the EAW resonance (boxed in a dash-dotted line) and the IAW feature (boxed in a dashed line). The spectrum at 9 ns (along the vertical dashed line of (b)) is plotted in (d) as the blue line with a fitted synthetic spectrum (red solid line). A two-stream electron distribution, shown as the solid red line in the inserted plot ($f_e$), is needed to reproduce the strong EAW signal. $-v$ direction is along the redshifted ${\bm k}$ in Fig.~\ref{fig:setup}b. The velocity at the valley of the distribution ($-0.023c$) matches the EAW’s phase velocity ($0.025c$). For reference, a Maxwellian distribution, shown as the dashed red line in the inserted figure, would generate the dashed red spectrum. 
}
    \label{fig:TSIAWEAWexpvssyn}
\end{figure*}

\begin{figure}[tb] 
    \centering
    \includegraphics{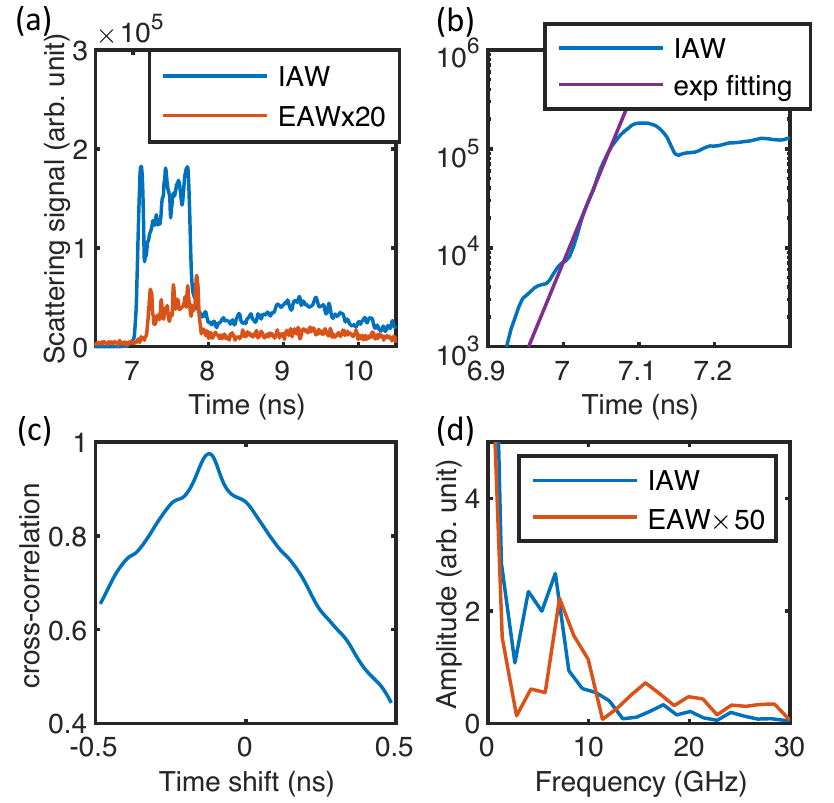} 
    \caption{Time evolution of IAW and EAW signals. (a) IAW and EAW Thomson scattering signal in the dash-boxed (IAW) dash-dot-boxed (EAW) region shown in Fig.~\ref{fig:TSIAWEAWexpvssyn}(b). Bursts of IAW and EAW are shown around 7 – 8 ns. 
    (b) is the zoomed-in plot of (a) in $\log_{10}$ scale to show the growth phase of the IAW signal, which is fitted by an exponential function with e-folding time 0.025~ns (3200 $\omega_{\rm pe}^{-1}$), agreeing with the IAW growth rate, $\gamma\sim 1.4\times10^{-4} \omega_{\rm pe}$, where $\omega_{\rm pe}$ is the electron plasma frequency.
    (c) Cross-correlation of the IAW and EAW signal shows a strong correlation between IAW and EAW. IAW is leading EAW by 0.12~ns$~\sim1.5\times10^4  \omega_{\rm pe}^{-1}$. 
    (d) The Fourier transform of the 7--8~ns signal shows that both IAW and EAW signals are oscillating with an amplitude peak frequency at 7 GHz. This frequency matches the lower-hybrid frequency in a 20 T magnetic field. 
    }
    \label{fig:IAWEAWlineout}
\end{figure}

The Thomson scattering from the ion-acoustic waves (IAW) reveals that 
current-driven instabilities develop at 7~ns.
The narrowband spectrometer captured the asymmetric (10:1) IAW Stokes and anti-Stokes peaks. The scattering signals grow from thermal level by three orders of magnitude to extremely intense, bursty, and asymmetric ($\sim$100:1) IAW peaks during 7-8~ns.
This is shown in Fig.~\ref{fig:TSIAWEAWexpvssyn}(a), as a sign of the ion turbulence induced by unstable IAW~\cite{Daughney1970}. 
The asymmetry in the IAW Stokes and anti-Stokes peaks is the feature of the drift between electrons and ions which differs the electron Landau damping rates for IAWs in two directions~\cite{Hawreliak2004, TSbook}.
As shown in Fig.~\ref{fig:TSIAWEAWexpvssyn}(c), the Thomson scattering spectrum lineout before the IAW bursts (along the dashed line in Fig.~\ref{fig:TSIAWEAWexpvssyn}a) can be reproduced in the synthetic spectrum (green dashed line in Fig.~\ref{fig:TSIAWEAWexpvssyn}c) when electrons ($n_{\rm e}=5\times10^{18}~{\rm cm^{-3}}, T_{\rm e}=200$~eV) are drifting with $v_{\rm d}=0.17 \sqrt{T_{\rm e}/m_{\rm e}}$ relative to ions ($T_{\rm i}=400~{\rm eV}, Z=18$).
This electron outflow speed is $\sim$5$v_{\rm A}$ and  $\sim$0.06$v_{\rm Ae}$.
The IAW-resonant peak is broader than the synthetic spectrum (green dashed line) that assumes flow velocity is uniform in the scattering volume, but the broader peak can be reproduced by including inhomogeneity of the flow velocity with $\Delta v=2\times 10^{4}$~m/s, which may be induced by a spatial gradient of the flow velocity or turbulence.
The synthetic spectrum calculation assumes the resonant wave is stable and the density fluctuations are at the thermal level. 
The scattering signals are at the thermal level at the initial stage before the IAW bursts.
However, based on the electrostatic dispersion equation, 
\begin{equation}
    1-\frac{\omega^2_{\rm pe}}{2 k^2T_{\rm e}/m_{\rm e}}{\rm Z}'\left(\frac{\omega/k-v_{\rm d}}{\sqrt{2T_{\rm e}/m_{\rm e}}}\right) \\-
    \frac{\omega^2_{\rm pi}}{2 k^2 T_{\rm i}/m_{\rm i}} {\rm Z}'\left(\frac{\omega /k}{\sqrt{2 T_{\rm i}/m_{\rm i}}}\right)=0,
\end{equation}
where ${\rm Z'}$ is the derivative of the plasma dispersion function~\cite{fried61}, 
the plasma with this strong electron drift is unstable to IAW, and the maximum growth rate is 17~ns$^{-1}$ ($1.4\times 10^{-4}\omega_{\rm pe}$, and 0.058~ns e-folding time) at $k=0.33/\lambda_{\rm De}$ ($\lambda_{\rm De}=\sqrt{T_{\rm e}/m_{\rm e}}/\omega_{\rm pe}$). 
The wavelength is 6 times shorter than the electrons' mean free path, which suggests that collisions are unimportant for the IAW growth.
This theoretical IAW growth rate agrees with the exponential growth of the scattering signal, which is proportional to the square of the density fluctuation $\delta n_{\rm e}^2(\omega, \bm{k})$. 
As shown in Fig.~\ref{fig:IAWEAWlineout}(b), during the growth of the first spike, the IAW signal rises exponentially with an 0.025~ns e-folding time or $3.3\times10^{-4}\omega_{\rm pe}$ growth rate, which is $\sim$2 times of the calculated IAW's growth rate. The agreement between the density fluctuation (scattering signal) growth rate and the calculated IAW growth rate also confirmed the electron drift speed.
In addition, the peak intensity of the IAW scattering is 3 orders higher than that when no burst presents near 5 ns, indicating that the fluctuation amplitude ($\delta n_{\rm e}(\omega, \bm{k})$ at $k=2\pi/527~{\rm nm}^{-1}$) is $\sim$30 times higher than the thermal level.

\begin{figure}[h]
    \centering
    \includegraphics{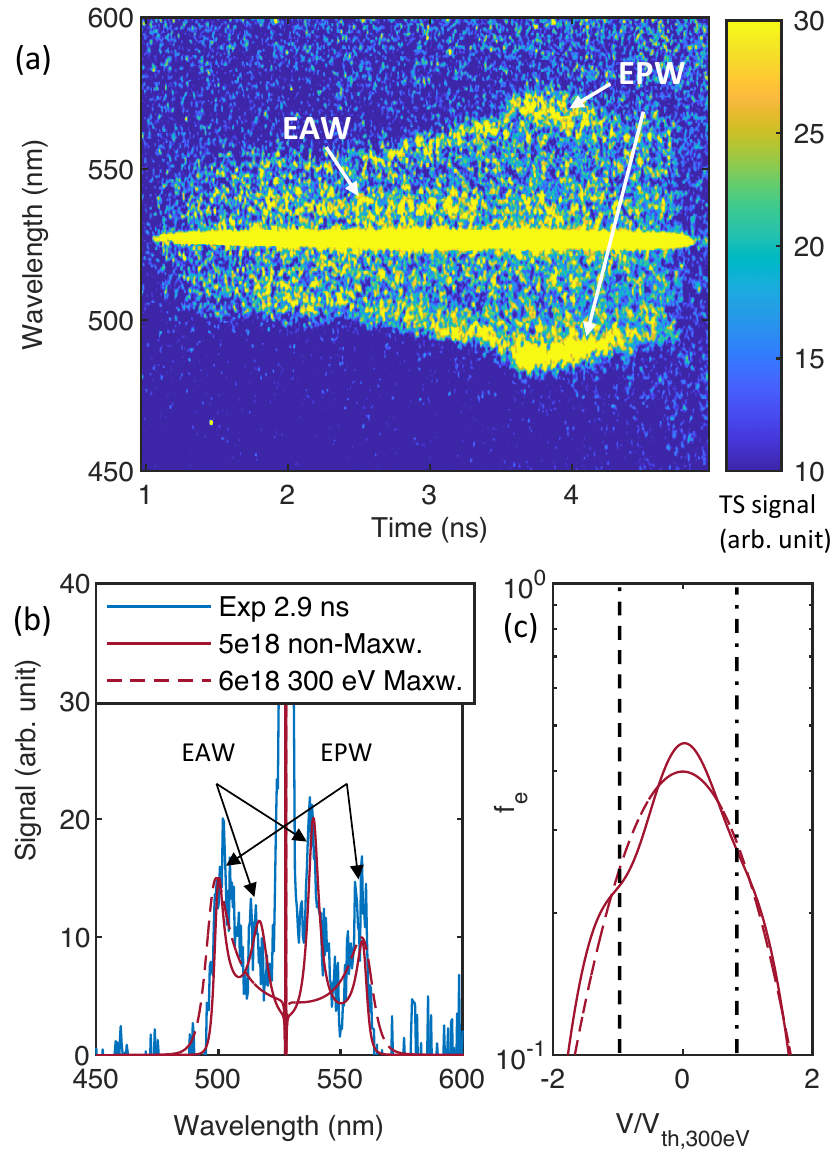} 
    \caption{Thomson Scattering data for EPW and EAW. (a) Time-resolved Thomson scattering (TS) shows features of electron plasma waves (EPW) and electron acoustic waves (EAW). The UV lasers onset at 0~ns. (b) Comparison between the measured TS spectrum at $t=2.9$~ns and the synthetic TS spectra with Maxwellian electrons (electron temperature $T_{\rm e}= 300$~eV, dashed) vs. non-Maxwellian electrons (solid). The velocity distribution functions $f_e$ are plotted in (c). The non-Maxwellian distribution (solid line in c) is constructed with secondary components (electron density $n_{\rm e}=1.15\times10^{18}$~cm$^{-3}$, $T_{\rm e}=75$~eV) counter-streaming with $-1.2v_{\rm th,300eV}$ and $+1.1v_{\rm th,300eV}$ relative to a steady component ($n_{\rm e}=2.7\times10^{18}$~cm$^{-3}$, $T_{\rm e}=75$~eV). $v_{\rm th}$ is the electron thermal velocity and $v_{\rm th,300eV}$ is the velocity when the temperature is at 300~eV. $-v$ direction is along the redshifted ${\bm k}$ in Fig.~\ref{fig:setup}b. This non-Maxwellian distribution with counter-streaming secondary components is required to match the measured spectrum since it avoids Landau damping near the EAW phase velocities by reducing the velocity slope. The EAW phase velocities are marked with a dashed line corresponding to the EAW at 539~nm and a dash-dotted line for the EAW at 517 nm. 
    }
    \label{fig:nonMaxw}
\end{figure}

With about 0.12 ns delay from the bursts of the IAW (see Fig.~\ref{fig:IAWEAWlineout}(c)), a scattering peak appears with a $\sim$13~nm red-shifted wavelength,  corresponding to a phase velocity of $0.025c\sim1.2 \sqrt{T_{\rm e}/m_{\rm e}}$, which is close to the electron thermal speed and matches the EAW's phase velocity. EAW scattering peaks are also observed in the earlier stage of the reconnection as shown in Fig.~\ref{fig:nonMaxw}. The appearance of the EAW scattering peak requires a non-Maxwellian velocity distribution with a flat or positive slope near the thermal speed to avoid Landau damping or stimulate waves.
The red solid line in the inserted figure of Fig.~\ref{fig:TSIAWEAWexpvssyn}(d) is the two-stream distribution that produces the synthetic spectrum with a strong EAW peak (red solid line). The velocity at the valley of the distribution matches the phase velocity of the EAW peak. 
The spectrum may allow other distributions to fit. Here we choose a two-stream distribution to reduce the complexity of the distribution function. 

The amplitude of IAW and EAW during the bursty period (7--8~ns) is shown to be modulated at a frequency of $\sim$7~GHz (Fig.~\ref{fig:IAWEAWlineout}d), close to the lower-hybrid frequency ($\sqrt{f_{\rm ce}f_{\rm ci}}$) at $B=20$~T.
One candidate to explain such observations is the Modified Two-Stream Instability (MTSI)~\cite{mcbride72, ji04} driven by the electron outflow jet perpendicular to the local magnetic field in the exhaust region.
The MTSI can generate electric field fluctuations nearly parallel to the current, which may modulate the IAW and the generation of EAW bursts. 
Waves near the lower-hybrid frequency were often observed in the MMS (Magnetospheric Multi-Scale) mission~\cite[e.g.][]{khotyaintsev20} and the MRX (Magnetic Reconnection Experiment)~\cite[e.g.][]{ji04} and have been suggested to mediate energy dissipation.
This modulation near the lower-hybrid frequency suggests instabilities like MTSI may affect the electron outflow, but further study is needed to characterize the role of this lower-hybrid modulation.

Electron heating is also captured since electron temperature increases by 60\% during the bursts of IAW and EAW. Electron temperature is measured from the separation between the IAW's Stokes and anti-Stokes peaks, which is proportional to the ion-acoustic velocity as $\Delta \omega\sim 2k \sqrt{ZT_{\rm e}/m_{\rm i}}$. 
Compared with the wavelength separation between the IAW peaks before the wave bursts (7.0~ns), this separation is 25\% larger after the IAW and EAW bursts (8.5~ns).

\section*{1D local particle-in-cell simulation}

\begin{figure}[h]
    \centering
    \includegraphics[width=8.5cm]{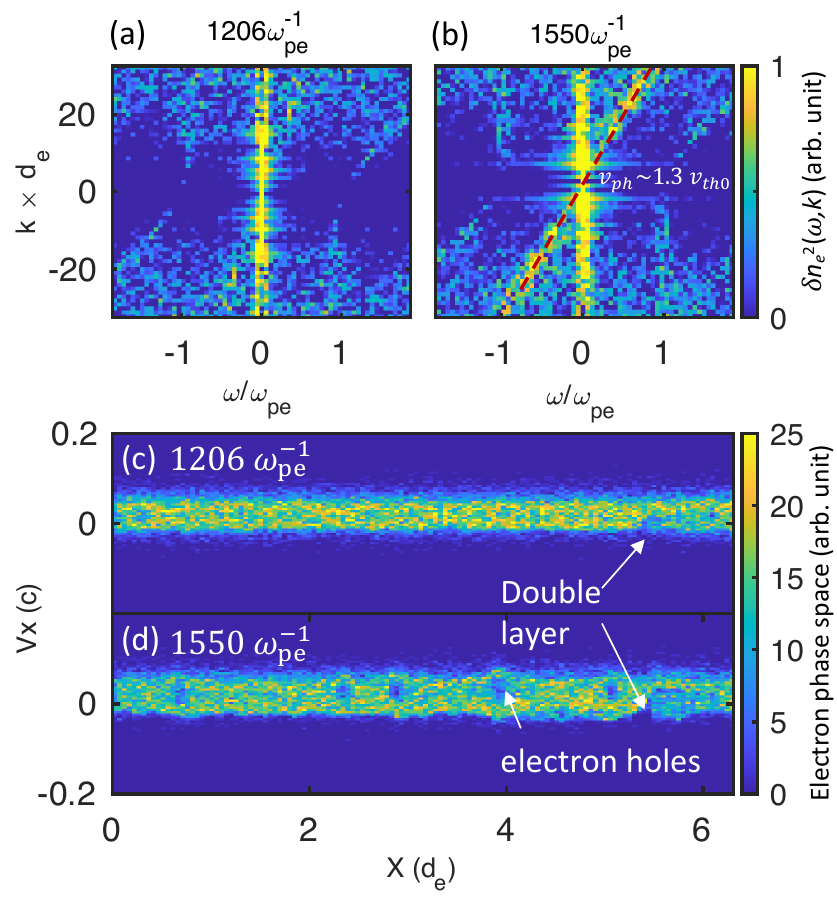} 
    \caption{Reproduction of IAW and EAW by 1D PIC simulation. (a) and (b) Electron density dispersion plots before ($t=1206\omega_{\rm pe}^{-1}$) and during the EAW bursts ($t=1550\omega_{\rm pe}^{-1}$), where $\omega_{\rm pe}$ is the electron plasma frequency. Wavenumber $k$ is normalized by $1/d_{\rm e}$ (electron skin depth $d_{\rm e}=c/\omega_{\rm pe}$). The red dashed line is the dispersion relation with phase velocity $v_{\rm ph}=1.3\sqrt{T_{\rm e}/m_{\rm e}}$, where $T_{\rm e}$ is the electron temperature, and $m_{\rm e}$ is the electron mass. (c) and (d) are the electron densities in $x-v_{x}$ phase space at the times of (a) and (b), respectively. (c) shows the double layer starts to form at $t=1206\omega_{\rm pe}^{-1}$. (d) After forming the double layer at $t=1550\omega_{\rm pe}^{-1}$, the electron holes are generated due to the two-stream instabilities downstream of the double layer. These electron holes are moving in the +x direction with a velocity of $1.3\sqrt{T_{\rm e}/m_{\rm e}}$, forming the EAW bursts shown in (b). }
    \label{fig:PIC}
\end{figure}

To understand the bursts of the correlated IAW and EAW, 
we have used a 1D electrostatic particle-in-cell code~\cite{Lapenta1DPIC} to simulate the thermal electrons (initial electron temperature $T_{\rm e0}=320$~eV) drifting relative to ions with a velocity $v_{\rm d}=0.5 \sqrt{T_{\rm e0}/m_{\rm e}}$, 
which is higher than the measured drift speed ($\sim$0.17$\sqrt{T_{\rm e0}/m_{\rm e}}$) to accelerate the process. In this simulation, the drifting electrons induce ion-acoustic instability and generate a double layer at $x\sim5d_{\rm e}$. This process agrees with the previous PIC~\cite{sato80} and Vlasov simulations~\cite{vazsonyi20}.

The bursts of IAW and EAW are reproduced in the 1D PIC simulation, as shown in Fig.~\ref{fig:PIC}. 
The dispersion plot Fig.~\ref{fig:PIC}(b) shows that the EAW burst has a phase velocity of $\sim$ $1.3 \sqrt{T_{\rm e0}/m_{\rm e}}$ with a broadband frequency of $\sim(0.1-1)~\omega_{\rm pe}$. This phase velocity roughly agrees with the experimentally observed EAW's phase velocity $\sim1.2\sqrt{T_{\rm e}/m_{\rm e}}$.
This EAW corresponds to the phase space holes shown in Fig.~\ref{fig:PIC}(d) since the holes are centered at $\sim$$\sqrt{T_{\rm e}/m_{\rm e}}$ and move forward with that speed. These electron holes originate from the electron-two-stream instability downstream of the double layer. This double layer reflects low-energy electrons and accelerates high-energy electrons that can overcome the potential well, resulting in a two-stream distribution, which has been discussed in previous Vlasov simulations~\cite{vazsonyi20}.
Their simulation also shows that, with a realistic mass ratio, the double layer occurs $\sim$ $10^4 \omega_{\rm pe}^{-1}$ after the peak of the IAW fluctuations. 
The delay between the IAW peak and the double layer generation is consistent with the observed 0.12~ns~$\sim1.5\times 10^{4}\omega_{\rm pe}^{-1}$ delay between the IAW and EAW bursts.

\section*{2D global particle-in-cell simulation}

\begin{figure}[h]
    \centering
    \includegraphics{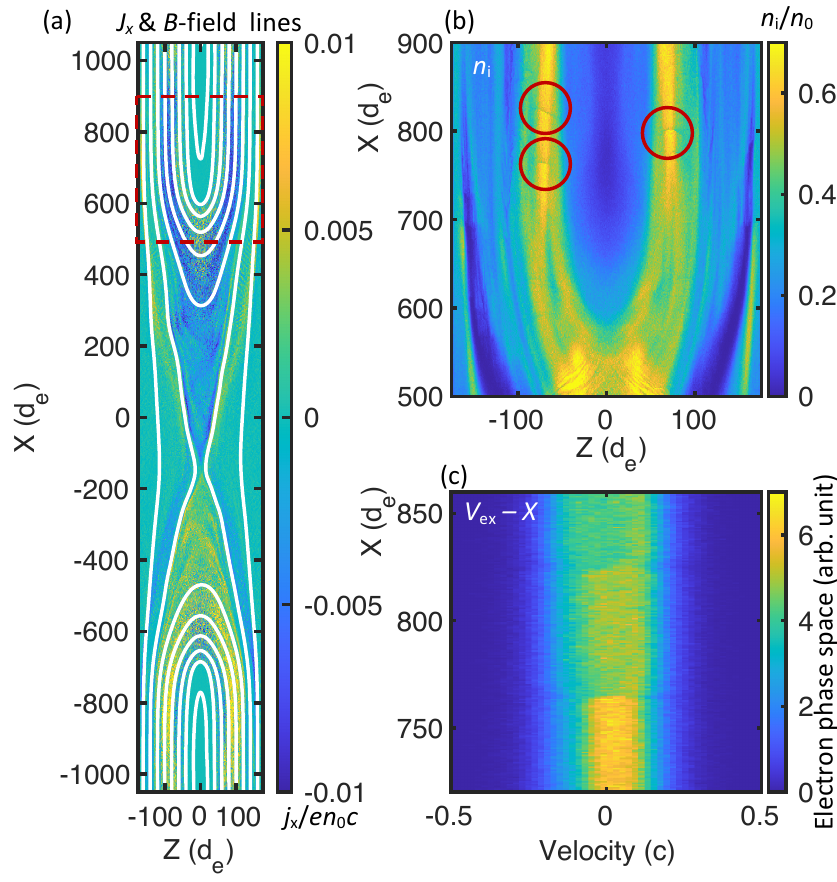} 
    \caption{Results of the 2D reconnection PIC simulation. The ion/electron mass ratio is $m_{\rm i}/m_{\rm e}=1600$ and time is 104000 $\omega_{\rm pe}^{-1}$. (a) The entire profile of the background electron current in $x$-direction ($J_x=e n_{\rm e,bg} v_x$ normalized to $e n_{\rm 0}c$) with magnetic field lines, where background electrons with density $n_{\rm e, bg}$ have an average velocity in $x$ direction ($v_x$). (b) A cropped region of the background plasma's ion density ($n_{\rm i}$, normalized by the peak density of the Harris current sheet $n_{\rm 0}$) in the outflow region (red dashes in a). The double layers, shown as the ion cavities, are circled in red. (c) Electron density profile in the phase space of x-direction velocity and x-axis ($v_{{\rm e}x}-x$) along $z=-74d_{\rm e}$ crossing the double layers at $x=765$ and 825 $d_{\rm e}$. $d_{\rm e}=c/\omega_{\rm pe}$ is the electron skin depth.}
    \label{fig:2DPIC}
\end{figure}

In addition to the 1D local PIC simulation showing the unstable IAW-generated double layer and EAWs, our 2D PIC reconnection simulation also confirms the double layer generation in the outflow region when cold background plasma is present. In the 2D reconnection simulation with cold background plasma, double layers in the outflow are developed and create non-Maxwellian and broadened distributions in the double layer downstream. Figure~\ref{fig:2DPIC}(a) is the in-plane current map and magnetic field lines in the entire simulation domain. 
The double layers are presented by the ion density cavities circled in red in the outflow region as shown in Fig.~\ref{fig:2DPIC}(b).
These density cavities can present double layers because the cavities coexist with ion phase space holes in this simulation, consistent with the Vlasov simulation~\cite{vazsonyi20}.
The electron phase space profile crossing the double layers at $z=-74 d_{\rm e}$ is plotted in Fig.~\ref{fig:2DPIC}(c). As shown in the phase space profile, upstream of the double layers, the electrons are drifting relative to ions, which can destabilize IAW forming double layers. In the downstream, the distribution is broadened and shows a non-Maxwellian distribution with double peaks. Besides the current-driven unstable IAW, ion--ion acoustic instability~\cite{gary87} is also shown in the region with two-streaming ions ($x\sim500 d_{\rm e}$ in Fig.~\ref{fig:2DPIC}b), especially in the 100 and 400 mass ratio cases. However, this ion--ion acoustic instability only creates strong density perturbations; no non-Maxwellian distribution is seen in electron phase space. Meanwhile, the double layers and the downstream non-Maxwellian distribution are persistent with different mass ratios. The observation of the current-driven double layers and the induced non-Maxwellian distribution confirm that, with cold background plasmas, the current-driven IAW bursts can result in energy dissipation in the outflow region. 
The amount of magnetic energy dissipated through this channel may depend on the plasma conditions such as $ZT_{\rm e}/T_{\rm i}$, $m_{\rm i}/m_{\rm e}$, plasma $\beta$, electron $\beta_{\rm e}$, and system size, which requires further comprehensive studies.

\section*{Discussion and outlook}

In summary, our low-$\beta$ 
magnetic reconnection experiments using laser-driven capacitor coils exhibit current-driven IAW bursts, followed by EAW bursts with electron heating in the exhaust region.

The location and wave direction are consistent with the IAWs observed by MMS~\cite{steinvall21}, THEMIS (Time History of Events and Macroscale Interactions during Substorms)~\cite{uchino17}, and PSP (Parker Solar Probe)~\cite{mozer2022} when a population of cold ions exists in the background, and their Landau damping is ineffective. 
The MMS observed IAWs in the outflow region with wavevectors in the direction along the $B$-field lines~\cite{steinvall21}, consistent with our observations. The PSP results show that triggered IAWs coincidence with the core electron heating~\cite{mozer2022}.
These observations suggest that the current-driven instabilities can lead to a bursty magnetic field energy release.
As our 1D and 2D PIC simulations reveal, this energy dissipation process involves IAW-formed double layers producing two-streaming electrons downstream, which induces the electron two-stream instability. This rapidly heats the electrons, braking the electron outflow jet in the ion-diffusion region. Such a double-layer-induced electron two-stream instability may also explain the origin of the EAWs observed by MMS~\cite{ergun16b}. 

This dissipation process in the reconnection exhaust region is confirmed in our experiment; whether this process can occur in the current sheet region needs further study. A similar mechanism has been observed in the current sheet of a 3D PIC simulation with a strong guide field, in which Buneman instability~\cite{buneman58} leads to a formation of double layers and triggers electron two-stream instability~\cite{che09}. In 1D simulations, the Buneman-instability-generated double layer also creates ion and electron phase space holes~\cite{smith82,goldman03}. However, without a guide field or with a weak guide field, the electrons would be deflected out of the current sheet within a short period, which is shown in 3D simulations with a finite guide field~\cite{daughton11}. The non-steady electron stream in the current sheet may interrupt the instabilities' growth. The growth of IAW and double layer needs $\sim$10$^{3}-10^{4} \omega_{\rm pe}^{-1}$, as suggested by our experiment. In addition to the time needed, Vlasov simulation and previous particle simulations~\cite{sato80} all demonstrate that generating the double layer requires a large system size ($>500 \lambda_{\rm De}$). Future experiments and large-scale 3D simulations are needed to study the current sheet region.

The IAW-type turbulence may be important for stellar flares and other plasmas where magnetic reconnection is prevalent, including those of black hole accretion engines. During the initial transient phase of stellar flares, for example, electrons are heated to high temperatures, and ions could remain cold and evade detection~\cite{polito18}. This condition favors destabilizing IAWs to dissipate current and thus magnetic free energy via electrostatic double layers, in turn triggering EAW and further heating electrons. Type-III and type-U radio emission~\cite{miteva07} could be generated by electron beams out of double-layer structures. 

Other low electron $\beta$ 
two-temperature plasmas, in which the electrons may be much cooler than ions, may exist in strongly magnetized black hole accretion disk corona \cite{dimatteo+1997} or regions within collisionless accretion flows \citep{narayan98,yuan14}. Here, the relative drift between electrons and ions can be sufficiently large compared with the electron thermal speed to overcome ion Landau damping due to simultaneous low density and low electron temperature (or equivalently electron $\beta \ll 1$), and thus unstable IAW or the related Buneman instabilities may be viable mechanisms to efficiently dissipate magnetic energy.  

In this context, we note that 
the observed current-driven unstable IAW provides a collisionless  coupling of ions and electrons: 
during the current-driven unstable IAW, the magnetic energy is converted to the ion energy in IAW and eventually forms the double layer that provides an electric potential. The double layer stores both ion kinetic energy and electric potential energy, which is then transferred to electron kinetic energy by accelerating electrons that pass through the double layer potential well, and heating electrons by the two-stream instability. In standard two-temperature accretion models used to explain curiously quiescent accretors, the rate of coupling between ions and electrons is assumed to be purely Coulomb collisional coupling, or parameterized \citep{narayan98,yuan14} freely.  In these models, the accretion produces low luminosity when the ions acquiring free energy from viscosity are unable to transfer their energy to radiating electrons on an accretion time scale. 
Whether a collisionless faster-than-Coulomb coupling exists in these contexts has been a long-standing open question because the answer can dramatically affect the paradigm as to why these sources appear so quiescent.
As such, it will be important to quantify how the specific mechanism that we have identified scales to the astrophysical contexts in future work. 

\section*{Acknowledgments}
The research presented is supported by the U.S. Department of Energy (DoE), Office of Science, Office of Fusion Energy Sciences High-Eenergy-Density Laboratory Plasma Science program under Award Number DE-SC0020103 (H.J., S.Z., A.C., L.G., E.B.).
The experiment was conducted at the Omega Laser Facility at the University of Rochester’s Laboratory for Laser Energetics with the beam time through the National Laser Users’ Facility (NLUF) Program supported by DoE/National Nuclear Security Administration (NNSA).
E.B. acknowledges support from DoE
grants DE-SC0020432 and DE-SC0020434,
and NSF grants 
AST-1813298 and PHY-2020249.
J.K, C.K.L., A.B., and R.P. are supported under the auspices of the U.S. DoE/NNSA under Contract DE-NA0003868. The FLASH code used in this work was in part developed by the DoE NNSA-ASC OASCR Flash Center at the University of Chicago. The authors would like to acknowledge the OSIRIS Consortium, consisting of UCLA and IST (Lisbon, Portugal) for providing access to the OSIRIS 4.0 framework supported by NSF ACI-1339893. We would like to thank Dr. Qing Wang, Dr. Lee Suttle, Dr. Jack Halliday, Dr. Sergey Lebedev, and Dr. William Daughton for fruitful discussions.

\section*{Author Contributions Statement}
H.J., L.G., and E.B. initiated the research.
S.Z., A.C., L.G., H.J., and E.B. designed the experiment with inputs from J.M., H.C., R.F., D.F., and J.K.
S.Z., R.F., D.F., and J.K. analyzed the Thomson scattering spectra. S.Z., A.C., L.G., H.J., J.K., C.K.L., A.B., and H.C. performed the experiments. 
C.K.L., A.B., and R.P. conducted and analyzed the proton radiography. 
H.J., E.B. contributed to the astrophysics implications. 
S.Z. performed the 1D and 2D PIC simulations, FLASH simulations, synthetic Thomson scattering simulations, and synthetic proton radiography simulations. 
S.Z., A.C., L.G., H.J., and E.B. contributed to the simulation data interpretations.  
S.Z., H.J., E.B., and L.G. wrote the manuscript. 
All authors read, revised, and approved the final version of the manuscript. 
\section*{Competing Interests Statement}

The authors declare no competing interests.


\input{main_final3.bbl}

\clearpage

\section*{Methods}

In the methods section, we present the setup of the Thomson scattering diagnostics, the calculation of the synthetic Thomson scattering spectrum, and the parameters of the 1D and 2D PIC simulations. The FLASH radiative-non-ideal magnetohydrodynamic simulations and the proton radiography used to confirm the existence of reconnection are presented in the supplemental material. 

\subsection*{Collective Thomson scattering}

In this experiment, a $f/10$ reflective collection system 63\degree{} off the probe's axis~\cite{katz12} collects the scattered light from a $60\times60\times50$~µm$^3$ volume near the focus. A narrowband (7 nm window) and a broadband (320 nm window) streaked spectrometers temporally and spectrally resolved the collected scattering light. The streak window is 5 ns. The narrowband spectrometer covers the light scattered by ion-acoustic waves. 
The broadband spectrometer can show the spectrum of the light scattered by EAW, EPW, and the merged IAW peaks.
The timing of the probe is changed for each shot to cover the entire reconnection process.

Collective Thomson scattering is used to diagnose the plasma conditions since the spectrum of the scattered light is sensitive to the plasma's density, temperatures, and velocities~\cite{TSbook}. In this experiment, the electron density and temperature are given by least-squares fitting the synthetic spectrum to the measured spectrum of the EPW resonances shown in Fig.~\ref{fig:nonMaxw}(a).  With the measured $T_{\rm e}$ and $n_{\rm e}$, we fit the IAW resonance peaks to diagnose the mean ionization level ($Z$) and ion temperature $T_{\rm i}$, because the separation between the IAW peaks is determined by the IAW's phase velocity ($\sqrt{ZT_{\rm e}/m_{\rm i}}$) and the width of the peaks and the peak-to-trough ratio are determined by $T_{\rm i}$. The $Z\sim18$ ionization level also agrees with the FLYCHK~\cite{FLYCHK} simulation. The shift of the IAW peaks is due to the Doppler shift, which gives the flow velocity. The asymmetry of the IAW peaks is used to calibrate the relative drift between electrons and ions since the drift can induce different electron Landau damping rates on the IAWs in two directions.
The plasma is collisionless for Thomson scattering since the electron mean free path is one order larger than the probe’s wavelength.

To forward fit the measured Thomson scattering spectrum, we calculated the synthetic power spectrum in Fig.~\ref{fig:TSIAWEAWexpvssyn} and Fig.~\ref{fig:nonMaxw} based on the model summarized in Ref.~\cite{TSbook}. 
The synthetic power spectrum for arbitrary velocity distributions is 
\begin{multline}\label{eq:powerspectrum}
    P(\lambda_s)\propto \left(1+\frac{2\omega}{\omega_{0}}\right)\left[\frac{2\pi}{k}\left\vert\frac{1+\chi_{\rm i}}{1+\chi_{\rm e}+\chi_{\rm i}}\right\vert^2 f_{\rm e}\left(\frac{\omega}{k}\right) \right.\\ \left. +\frac{2\pi Z}{k}\left\vert\frac{\chi_{\rm e}}{1+\chi_{\rm e}+\chi_{\rm i}}\right\vert^2f_{\rm i}\left(\frac{\omega}{k}\right)\right]\frac{d\omega}{d\lambda_s},
\end{multline}
where $\omega=\omega_{\rm s}-\omega_{0}$ is the angular frequency of the fluctuations scattering the probe light ($\omega_0, {\bm k}_0$) and generating the scattered light ($\omega_{\rm s}, {\bm k}_{\rm s}$), and $f_{\rm e,i}(v)$ are the electron/ion velocity distributions reduced to 1D along ${\bm k}$ direction. The electron and ion susceptibilities are given by
\begin{equation}
    \chi_{\rm e,i}(\omega, k)=\int_{-\infty}^{\infty}dv\frac{\omega_{\rm pe,i}^2}{k^2}\frac{k \partial f_{\rm e,i}/{\partial v }}{\omega-kv}. 
\end{equation}

\subsection*{Electron acoustic wave resonance in Thomson scattering}

Figure~\ref{fig:nonMaxw}(a) shows the time evolution of the Thomson scattering spectrum from the broadband spectrometer, and the spectrum at 2.9 ns is shown as the blue line in Fig.~\ref{fig:nonMaxw}(b). In addition to the Stokes and anti-Stokes scattering of the EPW usually seen in thermal plasmas, the Thomson scattering spectrum also shows resonant peaks with lower wavelength shift ($\sim$10~nm), indicating the non-Maxwellian distribution in the reconnection exhaust. 
These shorter wavelength resonant peaks are caused by EAWs with phase velocities near the electron thermal velocity $v_{\rm e,th}={\sqrt{T_{\rm e}/m_{\rm e}}}$, which would be Landau damped if the electron velocity distribution was Maxwellian, as shown in the red dashed line in Fig.~\ref{fig:nonMaxw}(b). To reproduce the EAW peaks, we modified the distribution function by combining two counter-streaming beams with the steady component to reduce the slope near the thermal speed shown as the solid red line in Fig.~\ref{fig:nonMaxw}(c). 
The solid line in Fig.~\ref{fig:nonMaxw}(b) shows the fitted scattering spectrum calculated based on Eq.~(\ref{eq:powerspectrum}). 
This three-component distribution is similar to the observed ring-core distribution in reconnection PIC simulations~\cite{shuster14}, in which the ring in the outflow-out-of-plane phase space ($v_x -v_y$) reduces to  two counter-streaming beams in $v_x$. This ring structure is likely  produced by the reconnected magnetic field ($B_y$) turning the accelerated electrons~\cite{bessho14}. Similar ring-core distribution has also been observed by MMS in the reconnection exhaust region~\cite{burch16,torbert18}. 

\subsection*{1D PIC simulation}
The simulation was performed in a $2\pi c/\omega_{\rm pe}$ ($2\pi d_{\rm e}$) periodic domain with a reduced ion mass $m_{\rm i}/m_{\rm e}=100$ and a lower ion temperature $T_{\rm i}=20$~eV to keep the ion thermal speed lower than the IAW's phase velocity. The simulation domain contains 256 cells and 128 particles per cell.
To sustain the electrons streaming in the simulation with limited size,
we have added an external electric field $E_x=10^{-5} m_{\rm e} c \omega_{\rm pe}/e$ to the electric field calculated by the Poisson equation for advancing the particles' velocity. 

\subsection*{2D PIC simulation}

The 2D PIC simulations use OSIRIS code~\cite{fonseca02,hemker2000thesis} to simulate a Harris current sheet~\cite{harris62} and a cold background plasma initialized with a density profile as
\begin{equation}
    n_{\rm bg}(z)=0.3 n_{\rm 0}\Big[\frac{1}{2}+\frac{1}{2}\tanh\Big(\frac{|z|-2L}{0.5L}\Big)\Big],
\end{equation}
where $n_0$ is the peak density of the Harris current sheet and $L=20d_{\rm e}$. This simulation setup is similar to the cold background simulation described in Ref.~\cite{norgren2021}. 
This setup allows a low ion temperature in the outflow region to avoid ion Landau damping for IAWs.
The Harris current sheet has hot ions with $T_{\rm i,harris}=5T_{\rm e,harris}$. The background plasma is initialized with $T_{\rm i,bg}=T_{\rm e,bg}=T_{\rm e, harris}/25$. 
The anti-parallel magnetic field is in $x$ direction with $B_x=B_0 \tanh(z/L)$, 
where $B_0/em_{\rm e}=\omega_{\rm ce}=0.5\omega_{\rm pe}$. A long-wavelength perturbation~\cite{birn01} with 0.01$B_{\rm 0}$ amplitude is included to initialize reconnection.
The simulation has a $2100d_{\rm e}\times 350 d_{\rm e}$ box size in $6144\times1024$ cells. 
The boundaries are periodic in the $x$ direction. The $z$ direction boundaries are reflective for particles and conductive for the electric field. The mass ratio has been scanned with $m_{\rm i}/m_{\rm e}=100,400,$ and 1600.

\section*{Data availability}
The experimental Thomson scattering spectrograms are available on request from corresponding authors. 
\section*{Code availability}
The synthetic Thomson scattering calculation code is available on request from the corresponding authors. 
The 1D electrostatic PIC simulation code is available in Ref.~\cite{Lapenta1DPIC}.
The Osiris 4.0 PIC simulation code is available to authorized users signed MoUs with the Osiris Consortium, consisting of IST and UCLA. FLASH rad-MHD code is available from flash.rochester.edu. 
The plasma dispersion relation calculation code is available on the PlasmaDispersionRelation repo at https://github.com/xiaoshulittletree. 


\clearpage

\section{Supplemental material}

\beginsupplement

\subsection{Radiative and non-ideal magnetohydrodynamic simulation using FLASH code for the laser-driven capacitor-coil target}

\begin{figure}[b]
    \centering
    \includegraphics[width=8.5cm]{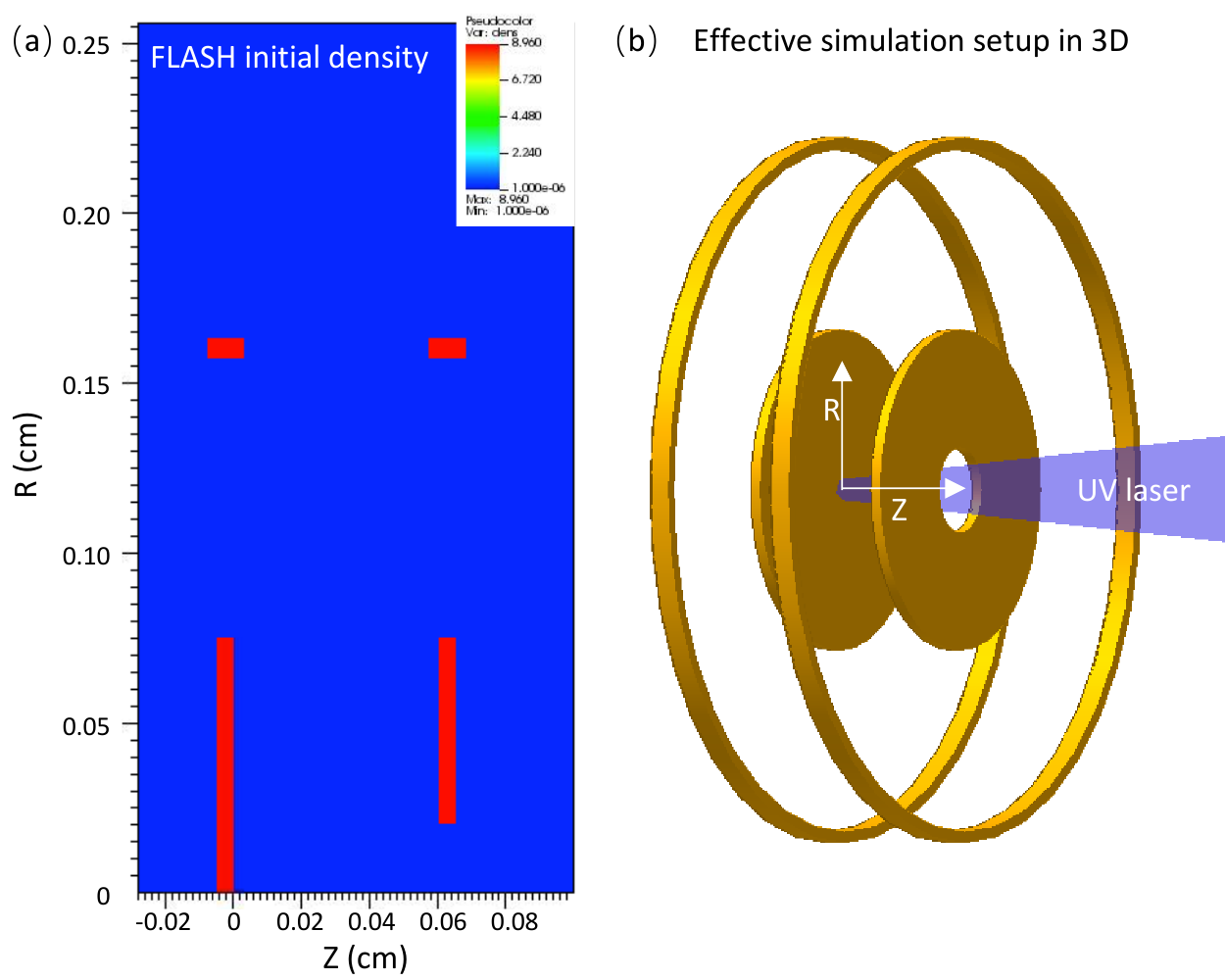}
    \caption{(a) Initial density profile of the FLASH simulation and (b) effective simulation setup in 3D. The two coils are located at $R=0.16$~cm and $Z=0$ and $0.06$~cm. Capacitor plates with a radius of 0.075 cm are centered at Z-axis at $Z=0$ and $Z=0.06$~cm. The 1-ns laser beam is along Z-axis. It irradiates the back plate at $Z=0$ through the entrance hole (200~\micro\meter{} radius) of the front plate at $Z=0.06$~cm.}
    \label{fig:FLASHinitial}
\end{figure}

To understand the magnetic field configuration, we have used the radiative and non-ideal magnetohydrodynamic (MHD) FLASH code to simulate the experiment. This simulation is simplified from the 3D geometry to 2D cylindrical geometry, in which the coils are represented by the rings in the planes of the plates. The initial density profile of this simulation is shown in Fig.~\ref{fig:FLASHinitial}(a). The simulation magnetic field configuration is benchmarked by the proton radiographs measured in the experiments. The benchmark is shown in the next section. 

The simulation reveals the plasma formed by the laser and the radiation. At $T=0$, a 1-ns UV laser with 100~\micro\meter{} focal spot is injected along $z$ axis, heating the back plate of the capacitor. The laser energy is reduced from 3~kJ total energy in the experiment to 2~kJ to compensate for the energy loss induced by the laser--plasma instabilities. The radiation from the laser spot also heats the coils and the top plate. The plasmas from the bottom and the top plates collide and squeeze out, generating plasma flowing toward the coils. 

In this FLASH simulation, the magnetic field is advanced based on the induction equation:
\begin{equation}
    \frac{\partial \mathbf{B}}{\partial t}=-\nabla
    \times \mathbf{E},
\end{equation}
where 
\begin{equation}
    \mathbf{E}= \mathbf{u}\times \mathbf{B}+\eta \mathbf{J} + \mathbf{E}_{\rm coil},
\end{equation}
$\mathbf{u}$ is the flow velocity, and $\mathbf{E}_{\rm coil}$ is the electric field in the coil due to the voltage between the capacitor plates. In the coil region where density $\rho>1~{\rm g/cm^{3}}$, $E_{\rm coil}=1.8\times10^{7}$~V/m in azimuthal direction. The corresponding voltage on a 2~mm coil is 36 kV. The resistivity function used in this simulation is
\begin{equation}\label{eq:eta}
\eta=\frac{T_{\rm e}}{\frac{5\times10^{6}}{8\times10^{9}} \frac{\rho}{8.96}+\frac{T_{\rm e}^{5/2}}{8.196\times10^{5} \bar{Z}\ln \Lambda}+\frac{3\times 10^{5}}{8\times10^{9}}\frac{\rho}{8.96}T_{\rm e}}.
\end{equation}
This resistivity is modified from Al's resistivity model~\cite{davies02}, 
\begin{equation}
\eta=\frac{1}{5\times10^{6} T_{\rm e}^{-1}+170T_{\rm e}^{3/2}+3\times 10^{5}} ~{\rm \Omega~m},
\end{equation} 
which is based on the measured Al resistivity from room temperature to $10^{6}$~K~\cite{milchberg88}.
In Eq.~(\ref{eq:eta}), $\eta$ is in FLASH's unit that $\frac{c^{2}}{4\pi}\eta_{\rm Gaussian} $=$\eta_{\rm FLASH}$. The density $\rho$ is in g/cm$^{-3}$, and the density factor ($\frac{\rho}{8.96}$) is to limit the non-Spitzer resistivity to the high-density coil region while keeping the low-density region as the Spitzer model.  
The initial coil temperature is set to 2900~K to reduce numerical instability. However, it would cause a higher-than-physical resistivity increasing the current dropping rate. Thus, we choose the duration of the voltage to be 3.5 ns, longer than the laser pulse length, to compensate for the higher coil resistivity induced by the high coil initial temperature.
The simulated coil current reaches 135 kA at 3.5 ns when the voltage is turned off. Then the current drops exponentially. The coil current is 59 kA at 6.0 ns, which is comparable to the $\sim$40--70~kA inferred from the proton radiography results.

\begin{figure}
    \centering
    \includegraphics[width=8.5cm]{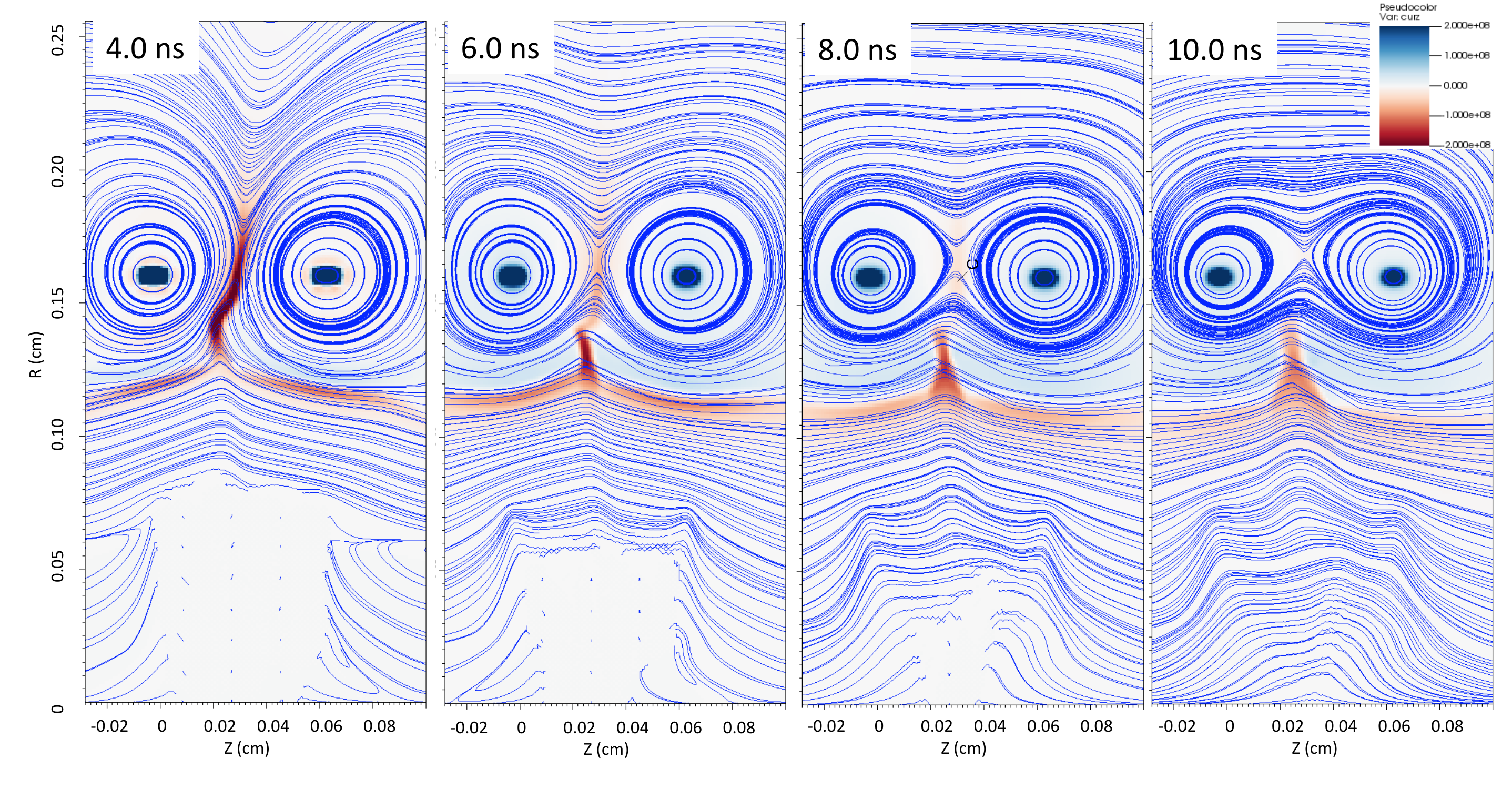}
    \caption{FLASH simulated out of plane current profile overlapped with magnetic field lines (blue lines). 
    }
    \label{fig:currentdensity}
\end{figure}

The simulation suggests that the current sheet is generated by the reconnection between the coils as well as the collision between the plate plasma and the coil plasmas. The current density profile and the magnetic field lines at 4~ns, 6~ns, 8~ns, and 10~ns are shown in Fig.~\ref{fig:currentdensity}. Compared with 4 ns, the current sheet in the center breaks into two parts at 6 ns. Then the reconnection current sheet drops to below $10^{7} ~{\rm A/cm^2}$ at 10 ns, while the collision-induced current sheet remains. 

The simulated magnetic field has been used to calculate the synthetic proton radiographs, and the synthetic radiographs are compared with the experimental ones. The synthetic proton radiographs show features similar to the experimental proton radiographs. This benchmark is discussed in the next section.

\subsection{Reconnection features in Proton radiography}\label{sec:Prad}

\begin{figure}[t] 
    \centering
    \includegraphics[width=8.5cm]{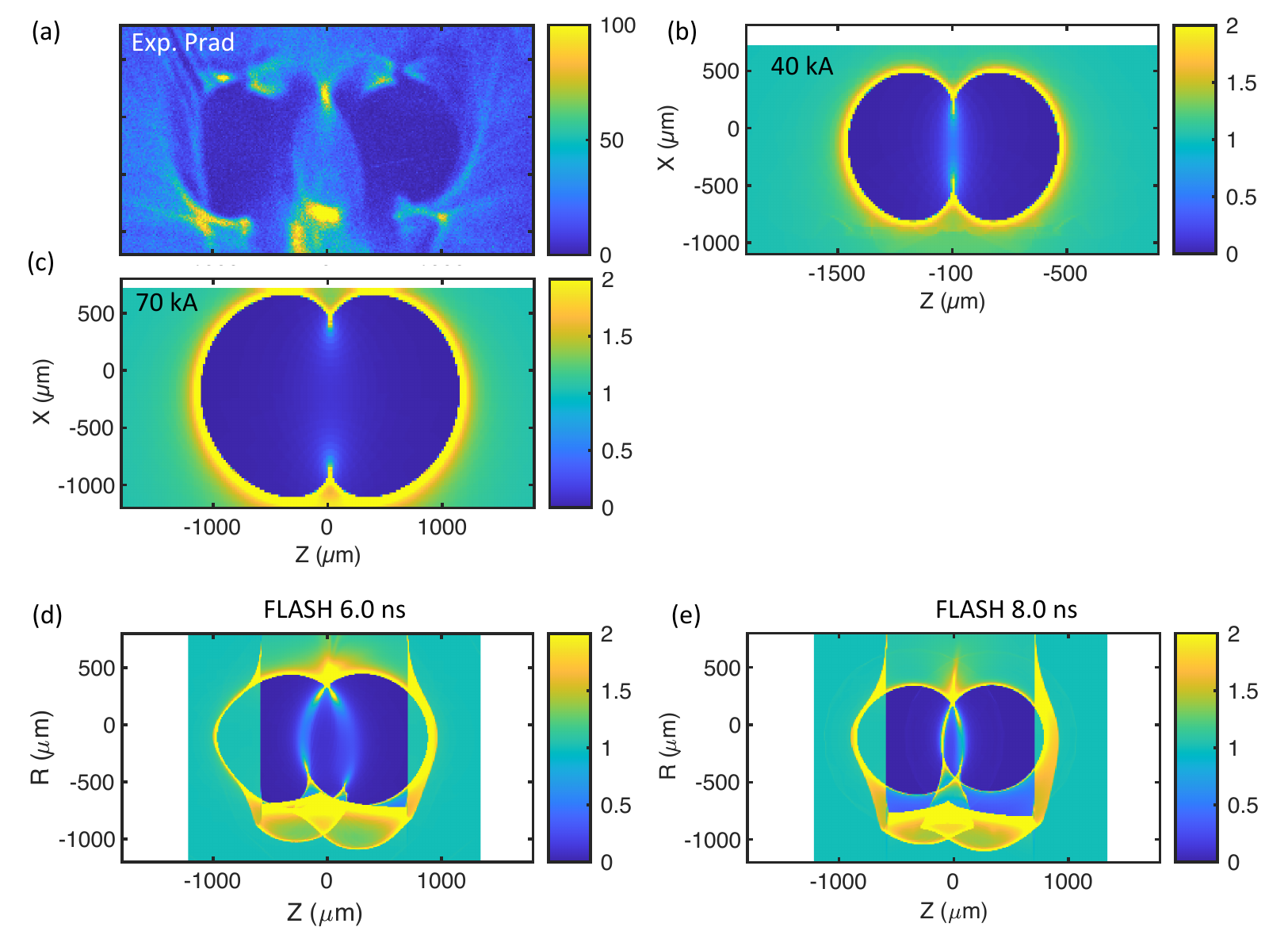}
    \caption{Experimental 15~MeV proton radiograph at 6.0 ns compared with synthetic proton radiographs (b, c, d, e). (b) Synthetic proton radiograph based on the Biot-Savart-law-calculated 3D magnetic profiles with 40 kA coil current. The vertical width of the two voids matches the measured total vertical width. (c) Synthetic proton radiograph with 70 kA coil current. The total horizontal width of the two voids matches the measured total horizontal void width. (d) and (e) are the synthetic proton radiograph based on the FLASH simulated B-field profiles. Both show a flask-like center feature similar to the one in the experiment, but the Biot-Savart-law-calculated proton radiographs did not show this feature. Coordinates in (d) and (e) have been shifted to locate the center between coils at $R=0, Z=0$. The B-field is filled artificially with 0 in the $|Z|>600~\micro\meter{}$ region out of the FLASH simulation domain, which resulted in uniform proton distribution in that outer region except the caustics deflected the coils.}
    \label{fig:FLASHproton}
\end{figure}

In the OMEGA experiment, monoenergetic protons generated by the implosion of a D$^3$He capsule backlit the capacitor-coil target to diagnose the magnetic field with proton radiography. The D$^3$He capsule is placed 10 mm in front of the capacitor-coil target, and the CR-39 detectors are placed 25~cm at the back of the target. Proton radiography has also been used in our OMEGA EP experiment with a similar capacitor-coil target as described in Ref.~\cite{gao16,chien19}; however, the protons in OMEGA EP are from Target Normal Sheath Acceleration (TNSA)~\cite{TNSASnavely2000} driven by the 700-fs pulse laser.

As shown in Fig.~\ref{fig:FLASHproton}(a), similar to the previous OMEGA EP experiment~\cite{chien19}, the OMEGA experiment also shows two voids near the coils and the flask-like feature between the voids. The voids can be reproduced in the synthetic proton radiograph with a Biot-Savart-law-calculated 3D B-field profile, as shown in Fig.~\ref{fig:FLASHproton}(b,c). They are due to the deflection of protons by the magnetic field around coils. However, without reconnection in the Biot-Savart-law calculation, the flask-like center feature cannot be reproduced. 

The flask-like feature can be generated by the pull-reconnection current or the Hall electric field, as discussed in Ref.~\cite{chien19}. However, this center feature can also be due to the push reconnection since it can be reproduced with the $B$-field profile of the FLASH simulation, in which the push reconnection lasts until 8 ns. 
Figure~\ref{fig:FLASHproton} (d) and (e) are the synthetic proton radiographs based on the FLASH simulated $B$-field profile. The center feature is present in the synthetic proton radiographs until 8~ns. However, the void and center feature sizes in the synthetic radiographs have some discrepancies with the measured radiograph. This discrepancy is expected since the simulation has a higher initial coil temperature, which results in a faster decay of the coil current and affects the reconnection. In addition, to calculate the synthetic proton radiographs with the 2D simulated B-field, we assume the B-field profile is uniform in a 0.3~mm thick region along the proton beam direction. The 3D structure of the B-field in the experiment may affect proton radiography. Also, since the simulation domain is smaller than the measurement, we need to assume no B-field out of the simulation domain. 

\begin{figure} 
    \centering
    \includegraphics[width=8.5cm]{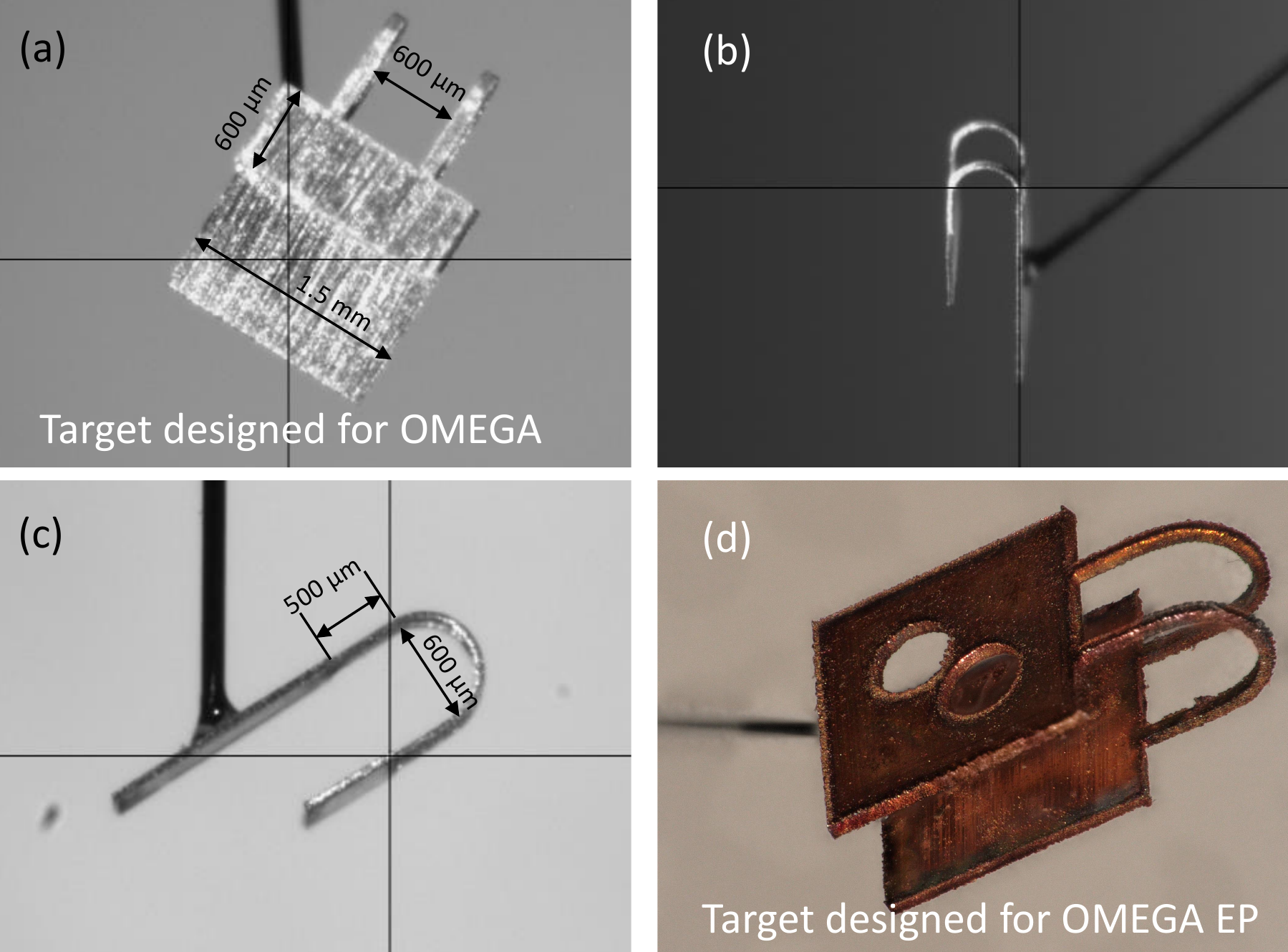}
    \caption{(a), (b), and (c) are the front view and side-on view of the target used in this OMEGA experiment. To better illustrate the geometry of the target, we also present (d) for the target used in the previous OMEGA EP experiment~\cite{chien19} with a similar design except for the front plate of the capacitor.}
    \label{fig:Targetpictures}
\end{figure}

In addition, even though the center feature may originate from both push and pull reconnection, the measured red-shifted IAW scattering confirmed an outflow in the scattering volume, which ruled out the possibility of the pull reconnection.

\end{document}

%% file: main_final3.bbl
%